\documentclass[twocolumn,showpacs,preprintnumbers,amsmath,amssymb]{revtex4}
\catcode`\@=11
\def\eqlt{\mathrel{\mathpalette\@vereq<}}  % < over =
\def\eqgt{\mathrel{\mathpalette\@vereq>}}  % > over =
\def\@vereq#1#2{\lower2.5pt\vbox{\baselineskip0pt \lineskip-.5pt
 \ialign{$\m@th#1\hfil##\hfil$\crcr#2\crcr{=}\crcr}}}

\newcommand{\simle}{\ \raise.3ex\hbox{$<$}\kern-0.8em\lower.7ex\hbox{$\sim$}\ }
\newcommand{\simge}{\ \raise.3ex\hbox{$>$}\kern-0.8em\lower.7ex\hbox{$\sim$}\ }

\usepackage{graphicx}% Include figure files
\begin{document}
%\draft
\title {Ferromagnetic fluctuation and possible triplet superconductivity in Na$_x$CoO$_2 \cdot y$H$_2$O: fluctuation-exchange study of multi-orbital Hubbard model}
\author {Masahito Mochizuki, Youichi Yanase and Masao Ogata}
\address {Department of Physics, University of Tokyo, Hongo Bunkyo-ku Tokyo 113-0033, Japan}
\date {today}

\begin{abstract}
Spin and charge fluctuations and superconductivity in a recently discovered superconductor Na$_x$CoO$_2 \cdot y$H$_2$O are studied based on a multi-orbital Hubbard model. Tight-binding parameters are determined to reproduce the LDA band dispersions with the Fermi surfaces, which consist of a large cylindrical one around the $\Gamma$-point and six hole pockets near the K-points. By applying the fluctuation-exchange (FLEX) approximation, we show that the Hund's-rule coupling between the Co $t_{2g}$ orbitals causes ferromagnetic (FM) spin fluctuation. Triplet $f_{y(y^2-3x^2)}$-wave and $p$-wave pairings are favored by this FM fluctuation on the hole-pocket band. We propose that, in Na$_x$CoO$_2 \cdot y$H$_2$O, the Co $t_{2g}$ orbitals and inter-orbital Hund's-rule coupling play important roles on the triplet pairing, and this compound can be a first example of the triplet superconductor in which the orbital degrees of freedom play substantial roles.
\end{abstract}

\maketitle
%\sloppy \maketitle

Recently, the first Co-based oxide superconductor Na$_x$CoO$_2 \cdot y$H$_2$O with $T_c \sim$5 K was discovered by Takada $et$ $al.$~\cite{Takada03}. In this compound, edge shared CoO$_6$ octahedra form CoO$_2$ planes similar to the high $T_c$ cuprates with CuO$_2$ planes. However, the Co ions form a triangular lattice in the CoO$_2$ plane. 
%It is also remarkable that the enhancement of two dimensionality by water 
%intercalation induces superconductivity. 
Mechanisms and nature of the superconductivity are currently attracting great interest because of a possible realization of unconventional pairing or resonating-valence bond (RVB) state~\cite{Anderson73,Anderson87}. Although there are still some controversies, extensive Co NMR/NQR experiments have shown evidences for non-$s$-wave pairing~\cite{Kato03,Ishida03,Fujimoto03}. Various experiments on the non-hydrate compounds show characteristic behaviors of strongly correlated electron systems~\cite{YWang04,HBYang03} and in-plane ferromagnetic (FM) spin fluctuations~\cite{Boothroyd04,Bayrakci04}. In spite of these extensive studies, however, the pairing symmetry has not been determined.

%From theoretical viewpoints, this compound has several interesting features. 
%One is geometrical frustration due to the triangular lattice and another is 
%the degeneracy of Co $3d$ orbitals. Originally, the RVB state was proposed 
%in the Heisenberg model on a triangular lattice. Actually high-temperature 
%expansion study suggests that the carrier doping to the triangular Heisenberg 
%model induces RVB superconductivity~\cite{Koretsune03}. Recent mean-field 
%analyses of $t$-$J$ model have shown that chiral $d_{xy}+id_{x^2+y^2}$-wave 
%pairing is stabilized~\cite{Baskaran03,Kumar03,QWang03,Ogata03}. However, the 
%broken time-reversal symmetry has not been observed so far~\cite{Higemoto04}, 
%and alternative states have been investigated.

According to the triangular crystal symmetry, superconducting states are classified as $s$-, $p$-, $d$-, $f_{x(x^2-3y^2)}$ (or next-nearest-neighbor $f$-), $f_{y(y^2-3x^2)}$, and $i$-wave pairings. Previous microscopic theories using various model Hamiltonians have shown that various symmetries arise depending on the model. For example, if one uses an extended Hubbard model which has a charge-ordering instability in large-$V$ region~\cite{Baskaran03b,Motrunich03}, one obtains the next-nearest-neighbor $f$-wave pairing~\cite{YTanaka04}. Mean-field analyses of $t$-$J$ model predict the chiral $d_{xy}+id_{x^2+y^2}$-wave pairing~\cite{Baskaran03,Ogata03,Kumar03,QWang03,Koretsune03}, and $d_{xy}+id_{x^2+y^2}$- and $f$-wave states are obtained in the single-band Hubbard model~\cite{Kuroki04,Ikeda04,Nisikawa04}. We think that this variety of superconducting states is due to frustration which does not give any specific momentum dependence of $\chi_s(q)$.

The above theories, however, assumed single-band models. As shown by LDA calculations~\cite{Singh00,Johannes04}, there is apparently an orbital degeneracy in Na$_x$CoO$_2$ systems. Actually two bands constructed from the three Co $t_{2g}$ orbitals intersect the Fermi level. Furthermore, the electron fillings of these two bands are far from half filling. This means that multi bands (in other words, multi orbitals) contribute to low-energy electronic structure. This is in marked contrast with the high-$T_c$ cuprates.

In this letter, we study the electronic structure and superconductivity in Na$_x$CoO$_2 \cdot y$H$_2$O on the basis of the multi-orbital Hubbard model, which includes the Co $t_{2g}$ orbitals and the inter- and intra-orbital interactions. We find that several important and interesting aspects appear which were not expected in the single-band model. One of them is a FM spin fluctuation which appears from the six hole pockets consisting of ${e'}_g$ orbitals as described below, and which is enhanced by the inter-orbital Hund's-rule coupling. Actually, an evidence for the FM fluctuation has been observed by Co-NQR experiments in several groups~\cite{Kobayashi03a,Kato03,Ishida03} and expected in the LSDA calculation~\cite{Singh03}. We will show that this FM spin fluctuation leads to triplet pairing mainly on the hole pockets. Here, the disconnectivity of the Fermi surface plays an important role. First, we will deduce the tight-binding parameters which reproduce the LDA band dispersions~\cite{Singh00,Johannes04}. Then we apply the fluctuation-exchange (FLEX) approximation to this multi-orbital Hubbard model. Generally speaking, the FLEX calculation is appropriate in the case near a quantum critical point because the critical enhancement of fluctuation can be taken into account. We point out that Na$_x$CoO$_2 \cdot y$H$_2$O can provide a first example of the triplet superconductivity which clarifies the important roles of orbitals on the superconductivity in the strongly correlated electron systems.

We study the multi-orbital Hamiltonian given by $H=H_{\rm cry.}+H_{\rm kin.}+H_{\rm int.}$ with $H_{\rm cry.}=\sum_{i,m,n,\sigma}D_{mn}d^{\dagger}_{im\sigma}d_{in\sigma}$, $H_{\rm kin.}=\sum_{i,j,m,n,\sigma} t_{ij}^{mn}d^{\dagger}_{im\sigma} d_{jn\sigma}-\mu\sum_{i,m,\sigma} d^{\dagger}_{im\sigma} d_{im\sigma}$, and $H_{\rm int.}=H_{U}+H_{U'}+H_{J_{\rm H}}+H_{J'}$, where $d_{im\sigma}$ ($d^{\dagger}_{im\sigma}$) is the annihilation (creation) operator of an electron with spin $\sigma$(=$\uparrow,\downarrow$) in the orbital $m$ on the Co site $i$. Here, $m$ runs over the $xy$, $yz$, and $zx$ orbitals. The first term $H_{\rm cry.}$ expresses the crystal field from the O ions acting on the Co $t_{2g}$ orbitals. In Na$_x$CoO$_2 \cdot y$H$_2$O, the CoO$_6$ octahedron is contracted along the $c$-axis. The ligand oxygens on the CoO$_6$ octahedron with this distortion generate a trigonal crystal field, and the matrix element of $D_{mn}$ is given by $\frac{\Delta}{3}(\delta_{mn}-1)$. This crystal field lifts the local $t_{2g}$ degeneracy into lower $a_{1g}$ and higher ${e'}_g$ levels with energy splitting of $\Delta$.

%%%%%%%%%%%%%%%%%%%%%%%%%%%%%%%%%%%%%%%%%%%%%%%%%%%%%%%%%%%%
\begin{figure}[tdp]
\includegraphics[scale=0.5]{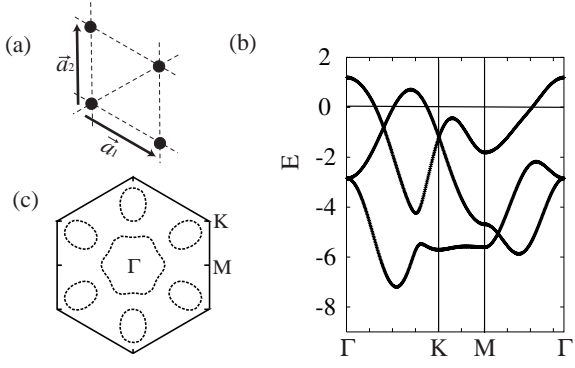}
%\epsfile{file=Fig01.eps,scale=0.5}
\caption{(a) Triangular lattice and its Bravis vectors. (b) Band dispersions and (c) Fermi surfaces obtained from $H_{\rm kin.}+H_{\rm cry.}$, which reproduce very well the characters of the LDA results~\cite{Singh00} especially near the Fermi level. The bands far below the Fermi level are relatively different from LDA results, implying a possible importance of further long-range hoppings and the O $2p$ components.}
\label{FermiS}
\end{figure}
%%%%%%%%%%%%%%%%%%%%%%%%%%%%%%%%%%%%%%%%%%%%%%%%%%%%%%%%%%%%%
The second term $H_{\rm kin.}$ represents the kinetic-energy. Here, $\mu$ is the chemical potential, and $t_{ij}^{mn}$ is the hopping integral between the $m$ orbital on the $i$ site and the $n$ orbital on the $j$ site. In order to reproduce the LDA~\cite{Singh00} band structures, we find that up to the third nearest-neighbor hoppings between orbitals are necessary. In the $k$-space, $H_{\rm kin.}$ is rewritten as
$H_{\rm kin.}=\sum_{{\bf k},m,n,\sigma}\epsilon^{mn}_{\bf k}
      d^{\dagger}_{{\bf k}m\sigma} d_{{\bf k}n\sigma}
     -\mu\sum_{{\bf k},m,\sigma}
      d^{\dagger}_{{\bf k}m\sigma} d_{{\bf k}m\sigma}$
with
%%%%%%%%%%%%%%%%%%%%%%%%%%%%%%%%%%%
\begin{eqnarray}
\epsilon^{\gamma\gamma}_{\bf k}&=&2t_1\cos k^{\gamma\gamma}_a + 2t_2\left[\cos k^{\gamma\gamma}_b + \cos (k^{\gamma\gamma}_a+k^{\gamma\gamma}_b)\right] \\ \nonumber
&+& 2t_4\left[\cos(2k^{\gamma\gamma}_a+k^{\gamma\gamma}_b) + \cos(k^{\gamma\gamma}_a-k^{\gamma\gamma}_b)\right] \\ \nonumber
&+& 2t_5\cos 2k^{\gamma\gamma}_a , \\
\epsilon^{\gamma\gamma'}_{\bf k}&=&2t_3\cos k^{\gamma\gamma'}_b + 2t_6\cos 2k^{\gamma\gamma'}_b \\ \nonumber
&+& 2t_7 \cos(k^{\gamma\gamma'}_a+2k^{\gamma\gamma'}_b) \\ \nonumber
&+& 2t_8 \cos(k^{\gamma\gamma'}_a-k^{\gamma\gamma'}_b) + 2t_9 \cos(2k^{\gamma\gamma'}_a+k^{\gamma\gamma'}_b).
\end{eqnarray}
%%%%%%%%%%%%%%%%%%%%%%%%%%%%%%%%%%%
Here, $\gamma$ and $\gamma'$ represent $xy$, $yz$ and $zx$ orbitals and 
$k^{xy,xy}_a=k^{xy,zx}_a=k_1$, $k^{xy,xy}_b=k^{xy,zx}_b=k_2$, 
$k^{yz,yz}_a=k^{xy,yz}_a=k_2$, $k^{yz,yz}_b=k^{xy,yz}_b=-(k_1+k_2)$, 
$k^{zx,zx}_a=k^{yz,zx}_a=-(k_1+k_2)$ and $k^{zx,zx}_b=k^{yz,zx}_b=k_1$, 
respectively. $k_1$ and $k_2$ are the components of the wave vector $\vec k$ of the triangular lattice spanned by $\vec {a_1}$ and $\vec {a_2}$, respectively (see Fig.~\ref{FermiS} (a)). In Fig.~\ref{FermiS}, we show the band dispersions (b) and the Fermi surfaces (c) obtained from $H_{\rm kin.}+H_{\rm cry.}$ in the case of $t_1=0.45$, $t_2=0.05$, $t_3=1$, $t_4=0.2$, $t_5=-0.15$, $t_6=-0.05$, $t_7=0.12$, $t_8=0.12$, $t_9=-0.45$ and $\Delta=0.4$. Both reproduce well the LDA results~\cite{Singh00}, particularly near the Fermi level. In the following, we use these parameters and $t_3=1$ as the energy unit.

As shown in Fig.~\ref{FermiS}, the Fermi surfaces consist of a large cylindrical one around the $\Gamma$-point and six hole pockets near the K-points. The cylindrical one has a dominant $a_{1g}$-orbital character, while the six hole pockets have an ${e'}_g$-orbital character. Koshibae and Maekawa~\cite{Koshibae03} discussed hidden Kagome lattices using the second-nearest neighbor hoppings. However, in their case, the obtained Fermi surfaces have six large hole pockets around the M-points in contrast to the LDA result. In this compound, the hybridizations between the neighboring Co $t_{2g}$ and $e_g$ orbitals are relatively strong because of the lattice structure with edge-shared CoO$_6$ octahedra, and a hopping process via the unoccupied $e_g$ orbitals aids the third nearest neighbor hopping but not the second nearest one. Thus, it is essential to take into account up to the third nearest-neighbor hoppings for reproducing the LDA result. In Fig.~\ref{FermiS} (b), the bands far below the Fermi level are relatively different from LDA results, implying a possible importance of further long-range hoppings and the O $2p$ components. However, the following results do not depend on these details far below the Fermi level.

The last term in $H$, $H_{\rm int.}$ represents the on-site $d$-$d$ Coulomb interactions, where $H_{U}$ and $H_{U'}$ are the intra- and inter-orbital Coulomb interactions, respectively, and $H_{J_{\rm H}}$ and $H_{J'}$ are the Hund's-rule coupling and the pair-hopping interactions, respectively. These interactions are expressed using Kanamori parameters, $U$, $U'$, $J_{\rm H}$ and $J'$, which satisfy the relations; $U'=U-2J_{\rm H}$ and $J_{\rm H}=J'$. The value of $U$ has been estimated as 3-5.5 eV by the photoemission spectroscopy~\cite{Chainani03} and as $\sim$3.7 eV by the $ab$-$initio$ calculation~\cite{Johannes04c}. The value of $J_{\rm H}$ for the Co$^{3+}$ ion is 0.84 eV. Thus, the ratio $J_{\rm H}/U$, which gives the strength of Hund's-rule coupling, is 0.15-0.28 in this compound.

We analyze this multi-orbital Hubbard model by extending the FLEX approximation to the multi-orbital case~\cite{Takimoto04}. In the FLEX approximation, RPA-type bubble diagrams and ladder diagrams are taken into account. By determining self-consistently the renormalized Green's function and fluctuation exchange self-energy, the effects of mode-mode coupling between charge and spin fluctuations and quasi-particle life time due to damping are incorporated, which weaken the Stoner-type instability expected in the mean-field type analysis. In the present three-orbital case, the irreducible Green's function $\hat{G}$, the non-interacting Green's function $\hat{G}^{(0)}$ and the self-energy $\hat{\Sigma}$ are expressed in the 3$\times$3-matrix form corresponding to the $xy$, $yz$ and $zx$ orbitals. The irreducible susceptibility $\hat{\chi}^0$ and the effective interaction $\hat{V}$ have a 9$\times$9-matrix form. The Green's function satisfies the Dyson-Gor'kov equation ${\hat{G}(k)}^{-1}={{\hat{G}^{(0)}}(k)}^{-1} -\hat{\Sigma}(k)$ with
%%%%%%%%%%%%%%%%%%%%%%%%%%%%%%%%%%%%%%%%%%%%%%%%%%%%%%%
\begin{equation}
 \Sigma_{mn}(k)=\frac{T}{N}\sum_{q}\sum_{\mu\nu} V_{\mu m,\nu n}(q)G_{\mu\nu}(k-q).
\end{equation}
%%%%%%%%%%%%%%%%%%%%%%%%%%%%%%%%%%%%%%%%%%%%%%%%%%%%%%%
Here, the matrix element of the effective interaction is given by
%%%%%%%%%%%%%%%%%%%%%%%%%%%%%%%%%%%%%%%%%%%%%%%%%%%%%%%
\begin{eqnarray}
 \!\!\!&&V_{mn,\mu\nu}(q) =[\frac{3}{2}\hat{U}^{\rm s}\hat{\chi}^{\rm s}(q)\hat{U}^{\rm s}
  +\frac{1}{2}\hat{U}^{\rm c}\hat{\chi}^{\rm c}(q)\hat{U}^{\rm c} \nonumber\\
 \!\!\!&&-\frac{1}{4}(\hat{U}^{\rm s} + \hat{U}^{\rm c})
   \hat{\chi}^0(q)(\hat{U}^{\rm s} + \hat{U}^{\rm c})
  +\frac{3}{2}\hat{U}^{\rm s}-\frac{1}{2}\hat{U}^{\rm c}]_{mn,\mu\nu}
\end{eqnarray}
%%%%%%%%%%%%%%%%%%%%%%%%%%%%%%%%%%%%%%%%%%%%%%%%%%%%%%%
with $\hat{\chi}^{\rm s}(q)=[\hat{I}- \hat{\chi}^0(q)\hat{U}^{\rm s}]^{-1} \hat{\chi}^0(q)$ and $\hat{\chi}^{\rm c}(q)=[\hat{I}+ \hat{\chi}^0(q)\hat{U}^{\rm c}]^{-1} \hat{\chi}^0(q)$. The irreducible susceptibility is given by
%%%%%%%%%%%%%%%%%%%%%%%%%%%%%%%%%%%%%%%%%%%%%%%%%%%%%%%
\begin{equation}
  \chi^0_{mn,\mu\nu}(q) =-\frac{T}{N}\sum_{k}G_{\mu m}(k+q)G_{n\nu}(k).
\end{equation}
%%%%%%%%%%%%%%%%%%%%%%%%%%%%%%%%%%%%%%%%%%%%%%%%%%%%%%%
In the above, $T$ is temperature, $k \equiv ({\bf k},i\omega_n)$ and $q \equiv ({\bf q},i\nu_{l})$ with $\omega_n=(2n-1)\pi T$ and $\nu_l=2l\pi T$, $N$ is the number of sites, and $\hat{I}$ is a unit matrix. The interaction matrices $\hat{U}^{\rm s}$ and $\hat{U}^{\rm c}$ are represented as
%%%%%%%%%%%%%%%%%%%%%%%%%%%%%%%%%%%%%%%%%%%%%%%%%%%%%%%
\begin{eqnarray}
 \hat{U}^{\rm s}
 =\left[\begin{array}{cc}
   \hat{U^{\rm s}_1} & \hat{0} \\
   \hat{0}  & \hat{U^{\rm s}_2}\\
        \end{array}\right],
 \hat{U}^{\rm c}
 =\left[\begin{array}{cc}
   \hat{U^{\rm c}_1} & \hat{0} \\
   \hat{0}  & \hat{U^{\rm c}_2}\\
        \end{array}\right].
\end{eqnarray}
%%%%%%%%%%%%%%%%%%%%%%%%%%%%%%%%%%%%%%%%%%%%%%%%%%%%%%%
where $[U^{\rm s}_1]_{aa,bb}$ ($[U^{\rm c}_1]_{aa,bb}$) is $U$ ($U$) for $a=b$, and $J_{\rm H}$ ($2U'-J_{\rm H}$) for $a\ne b$. On the other hand, $[U^{\rm s}_2]_{ab,cd}$ ($[U^{\rm c}_2]_{ab,cd}$) with $a\ne b$ and $c\ne d$ is given by $U'$ ($-U'+2J_{\rm H}$) for $(a,b)=(c,d)$ and is given by $J'$ ($J'$) for $(a,b)=(d,c)$, and is 0 for the other cases.

To discuss the nature of superconductivity, we solve the following Eliashberg equation;
%%%%%%%%%%%%%%%%%%%%%%%%%%%%%%%%%%%%%%%%%%%%%%%%%%%
\begin{eqnarray}
    \lambda \phi_{mn}(k)&=&-\frac{T}{N}\sum_{q}\sum_{\alpha\beta}\sum_{\mu\nu}
    \Gamma^{\eta}_{\alpha m,n\beta}(q) \phi_{\mu\nu}(k-q) \nonumber\\
    &\times& G_{\alpha\mu}(k-q) \hspace{1mm} G_{\beta\nu}(q-k).
\end{eqnarray}
%%%%%%%%%%%%%%%%%%%%%%%%%%%%%%%%%%%%%%%%%%%%%%%%%%%
The singlet and triplet pairing interactions $\Gamma^{\rm s}$  and $\Gamma^{\rm t}$ have 9$\times$9-matrix forms as,
%%%%%%%%%%%%%%%%%%%%%%%%%%%%%%%%%%%%%%%%%%%%%%%%%%%
\begin{eqnarray}
 \!\!\!\!\!\!\hat{\Gamma}^{\rm s}(q)
   =\frac{3}{2}\hat{U}^{\rm s}\hat{\chi}^{\rm s}(q)\hat{U}^{\rm s}
    -\frac{1}{2}\hat{U}^{\rm c}\hat{\chi}^{\rm c}(q)\hat{U}^{\rm c}
    +\frac{1}{2}(\hat{U}^{\rm s}+\hat{U}^{\rm c}),\\
 \!\!\!\!\!\!\hat{\Gamma}^{\rm t}(q)
   =-\frac{1}{2}\hat{U}^{\rm s}\hat{\chi}^{\rm s}(q)\hat{U}^{\rm s}
     -\frac{1}{2}\hat{U}^{\rm c}\hat{\chi}^{\rm c}(q)\hat{U}^{\rm c}
    +\frac{1}{2}(\hat{U}^{\rm s}+\hat{U}^{\rm c}).
\end{eqnarray}
%%%%%%%%%%%%%%%%%%%%%%%%%%%%%%%%%%%%%%%%%%%%%%%%%%%
The eigenvalue $\lambda$ is a measure for the dominant superconducting symmetry, and becomes unity at $T=T_{\rm c}$. 

Calculations are numerically carried out with 32$\times$32 $k$-meshes in the first Brillouin zone, and 4096 Matsubara frequencies. The value of $U$ is fixed at 6.0, and temperature $T$ is fixed at 0.02.
%%%%%%%%%%%%%%%%%%%%%%%%%%%%%%%%%%%%%%%%%%%%%%%%%%%%%%%%%%%%
\begin{figure}[tdp]
\includegraphics[scale=0.4]{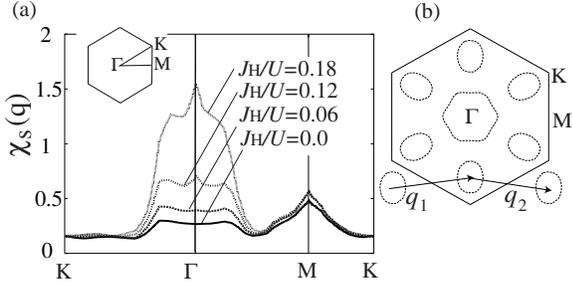}
%\epsfile{file=Fig02.eps,scale=0.4}
\caption{(a) Maximum eigenvalue of $\hat{\chi}^{\rm s}$ showing the evolution of FM fluctuation, as the strength of Hund's-rule coupling $J_{\rm H}/U$ increases. (b) Fermi surfaces for interacting case with $U$=6.0 and $J_{\rm H}/U$=0.18.}
\label{FMfluc}
\end{figure}
%%%%%%%%%%%%%%%%%%%%%%%%%%%%%%%%%%%%%%%%%%%%%%%%%%%%%%%%%%%%%
First, let us discuss the spin and charge fluctuations. Figure~\ref{FMfluc} (a) shows the maximum eigenvalue of $\hat{\chi}^{\rm s}$ obtained by solving the above FLEX equations self-consistently for various values of $J_{\rm H}/U$ at $T=$0.02. Apparently, a peak structure around $\Gamma$-point is strongly enhanced as $J_{\rm H}/U$ increases, while the small peak at the M-point hardly changes. Here, we also show the Fermi surfaces for interacting case with $U=$6.0 and  $J_{\rm H}/U$=0.18 obtained by the present FLEX calculation in Fig.~\ref{FMfluc} (b). The peak at the M-point originates from the inter-hole-pocket scatterings with wave numbers of $q_1$ and $q_2$ shown in this figure~\cite{Johannes04b}. Since the value of $J_{\rm H}/U$ is 0.15-0.28 in the actual compound, the enhanced peak at $\Gamma$-point at $J_{\rm H}/U=$0.18 indicates that the FM fluctuation is dominant in this compound. As mentioned before, FM fluctuation has been observed experimentally~\cite{Kobayashi03a,Kato03,Ishida03}. 
%Furthermore, the structure at the $\Gamma$-point is considerably small without Hund's-rule 
%coupling ($J_{\rm H}/U=0$), indicating a substantial importance of the Hund's-rule coupling 
%for the emergence of this FM fluctuation. 
On the other hand, $\hat{\chi}^{\rm c}$ does not exhibit any remarkable structures, which implies an absence of charge- and orbital-density-wave instabilities.

%%%%%%%%%%%%%%%%%%%%%%%%%%%%%%%%%%%%%%%%%%%%%%%%%%%%%%%%%%%%
\begin{figure}[tdp]
\includegraphics[scale=0.5]{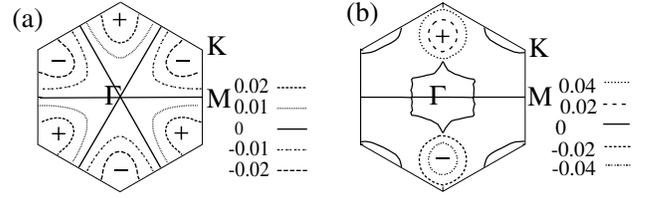}
%\epsfile{file=Fig03.eps,scale=0.5}
\caption{$k$-dependence of the superconducting gaps with the largest $\lambda$: (a) $f_{y(y^2-3x^2)}$-wave gap and (b) $p_y$-wave gap.}
\label{fwaveSC}
\end{figure}
%%%%%%%%%%%%%%%%%%%%%%%%%%%%%%%%%%%%%%%%%%%%%%%%%%%%%%%%%%%%%
By solving the Eliashberg equations, we find that triplet $f_{y(y^2-3x^2)}$-wave or $p$-wave pairings have the largest $\lambda$ in the calculated parameter region of $0<J_{\rm H}/U<0.30$. Although the values of $\lambda$ are nearly the same, $\lambda$ for $f_{y(y^2-3x^2)}$-wave state is slightly larger in the region of $J_{\rm H}/U<0.2$ while $\lambda$ for $p$-wave state is larger in $J_{\rm H}/U>0.2$. The other pairing instabilities are considerably weak. We have also studied a second-order perturbation theory and have obtained consistent results~\cite{Yanase04}.

Figure~\ref{fwaveSC} shows the $k$-dependence of obtained superconducting gaps with (a) $f_{y(y^2-3x^2)}$-wave and (b) $p_y$-wave symmetries. As for the $p$-wave, the $p_x$ and $p_y$ states are degenerate in the triangular lattice. In addition to it, there are three choices of the direction of $d$-vector ($\hat{\rm x}$,$\hat{\rm y}$,$\hat{\rm z}$), i.e., the $S_z$-component of $S$=1 Cooper pair. Below $T_c$, a linear combination of these six states should be realized, namely, $\hat{\rm z}(p_x \pm ip_y)$, $\hat{\rm x}p_x \pm \hat{\rm y}p_y$ and $\hat{\rm x}p_y \pm \hat{\rm y}p_x$. All of their gaps are represented as $\sqrt{\Delta_x(k)^2+\Delta_y(k)^2}$ with $\Delta_x(k)$ ($\Delta_y(k)$) being an order parameter for $p_x$ ($p_y$) state. This gap structure has line nodes on the $a_{1g}$ Fermi surface since the line nodes of $p_x$ and $p_y$ states have nearly six-fold symmetry as shown in Fig.~\ref{fwaveSC} (b). It is apparent that the magnitude of gap is large on the ${e'}_g$ band with six hole-pocket Fermi surfaces, while it is considerably small on the $a_{1g}$ band. 
%On the other hand, the amplitude of the gap on the $a_{1g}$ band with a large 
%cylindrical Fermi surface is considerably small. 
The geometry of the Fermi surfaces plays a very important role: Because of the disconnectivity of the Fermi surfaces, the nodal lines run between the ${e'}_g$ hole pockets and do not intersect them. Consequently, the gap is fully opened on each ${e'}_g$ Fermi surface, while $a_{1g}$ Fermi surface has line nodes. As discussed before the discovery of Na$_x$CoO$_2 \cdot y$H$_2$O, this kind of disconnectivity favors superconductivity by avoiding the disadvantage from nodes on the Fermi surface~\cite{Kuroki01a}. The dominant contribution of the ${e'}_g$ band to the superconductivity is also attributed to the van Hove singularity (vHS) in the ${e'}_g$ band. As can be seen in the band structure in Fig.~\ref{FermiS}, there exist saddle points near the K-points. This leads to a large density of states (DOS) near the Fermi level, which was actually reproduced in the LDA calculation~\cite{Singh00}.

Experimental estimate of the value of $J_{\rm H}/U$ has an ambiguity and is ranged from 0.15 to 0.28. Thus, both $f_{y(y^2-3x^2)}$ state ($J_{\rm H}/U<0.2$) and $p$ state ($J_{\rm H}/U>0.2$) can be candidates for the pairing state in the actual material. Now, let us discuss the present results with experiments. Both $f$-wave and $p$-wave gap structures obtained here are consistent with several NQR/NMR experiments~\cite{Fujimoto03,Ishida03}, which suggest existence of line nodes. Time-reversal symmetry observed in $\mu$SR~\cite{Higemoto04} excludes the possibility of chiral $p$-wave state $\hat{\rm z}(p_x \pm ip_y)$. Furthermore, the recent measurements of Knight shift~\cite{Kobayashi03a,Ishida04} and $H_{c2}$~\cite{FChou03,Sasaki04} are consistent if the $d$-vector is fixed to be parallel to the plane. This supports the $\hat{\rm x}p_x \pm \hat{\rm y}p_y$ or $\hat{\rm x}p_y \pm \hat{\rm y}p_x$ states. However, a contradictory result about the constant Knight shift below $T_c$ has also been reported~\cite{Kato03} so that further careful study is needed. In principle, the spin-orbit interaction lifts the degeneracy and determines which state is stabilized. This is an interesting future problem.

The existence of ${e'}_g$ hole pockets is still controversial. Several band calculations have predicted it~\cite{Singh00,Johannes04} while a recent LSDA+U calculation showed only a large $a_{1g}$ Fermi surface~\cite{PZhang05}. However, it was found that the L(S)DA+U cannot reproduce the experimental optical spectra~\cite{Johannes04c}, suggesting that this method is probably ill-suited for the present Co oxides. The angle-resolved photoemission spectroscopy has been performed only for the non-hydrate material~\cite{HBYang03} and the hole pockets have not been detected, which may be due to a surface effect or some matrix elements. Our calculation shows that the hole pockets play a substantial role for superconductivity. Indeed, if we study the situation without hole pockets by tuning the transfer parameters, we find that the pairing instability is strongly suppressed and the realization of superconductivity with $T_c\sim5$ K is difficult. Our results strongly suggest the existence of hole pockets.

In summary, we have studied the spin and charge fluctuations and the superconductivity in Na$_x$CoO$_2 \cdot$1.3H$_2$O on the basis of the FLEX analysis of the multi-orbital Hubbard model. The Hund's-rule coupling between the Co $t_{2g}$ orbitals gives rise to the FM spin fluctuation in agreement with the NQR results from several groups~\cite{Kobayashi03a,Kato03,Ishida03}. The  triplet pairing instability mediated by this FM fluctuation arises mainly on the ${e'}_g$ band with the pocket Fermi surfaces. The hole-pocket Fermi surfaces as well as a large DOS due to the vHS near the Fermi level are crucially important for the superconductivity. We have pointed out that this material can be a first example of the inter-orbital-interaction-induced triplet superconductivity. This mechanism is a universal one which can be expected in superconductors with multi-orbital degrees of freedom.

We thank K. Ishida, G.-q.Zheng, K. Yoshimura, Y. Kobayashi, M. Sato, R. Kadono, K. Kuroki, G. Baskaran for valuable discussions. M.M also thanks S. Onari and Yasuhiro Tanaka for discussions. This work is supported by ``Grant-in-Aid for Scientific Research'' from the MEXT of Japan and by NAREGI nanoscience project.
%%%%%%

\end{document}